\documentstyle[secnumtab]{fbssuppl}
\begin{document}

\title{Halo Structure Of $^{19}C$ Via The $(p,d)$ Reaction}
\author{B. G\"{o}n\"{u}l, and M. Yilmaz
\institute{Department of Engineering Physics, University of
Gaziantep,Gaziantep, 27310, Turkiye} \thanks{{\it E-mail
address':} gonul@gantep.edu.tr and ayilmaz@gantep.edu.tr}}
\maketitle

\begin{abstract}
We have investigated the heaviest one-neutron halo candidate $^{19}C$
nucleus. Few-body model calculations of cross section angular distributions
for the $^{19}C(p,d)^{18}C$ reaction, together with the test calculations
carried out for the $^{17}C(p,d)^{16}C$ reaction, at a low incident energy
are presented for different possible halo-neutron configurations. We show
that there is a clear distinction between in particular $\ell _n=0$ and $%
\ell _n=2$ halo transfers. The sensitivity of the cross sections to the
assumed $^{19}C$ single neutron separation energy is discussed.
\end{abstract}

\section{Introduction}

Dripline physics is at the focus of present-day nuclear science.
Experimentally, thanks to existing and emerging radioactive-ion-beam
facilities, we are on the verge of invading the territory of extreme $N/Z$
ratios in an unprecedented way. Theoretically, nuclear exotica represent a
formidable challange for the nuclear many-body theories and their power to
predict nuclear properties far from stability. In order to give a unified
description of all nuclei in the periodic table, theoretical systematic
studies in a single model are desired for both exotic nuclei near the
particle-drip line and ordinary nuclei near the stable line. Certainly, this
is a heavy task for future research.

A series of experiments \cite{Hansen} have shown that there exist neutron
halos in exotic light nuclei near the neutron-drip line. These studies
change our views on nuclear structure and lead to a new impact to the
traditional nuclear physics along the isospin freedom. Going to the limits
of the nuclear binding is also important for an improvement of our
description of normal nuclei from the neighborhood of the beta stability
valley. As our knowledge on nuclei far from stability deepens, it becomes
more apparent that nuclear halos are a general feature of loosely bound
nuclei. As the binding energy becomes smaller in the vicinity of the drip
lines, the valence nucleon(s) tunnel out of the central potential, and
enhance the diffuseness of the nuclear surface. Eventually, this leads to a
delocalization of the valence nucleon(s) which can be pictured as a halo
surrounding the core of the nucleus. The appearance of this halo is
determined by the height of the potential barrier, which itself depends on
the binding energy, the angular momentum and, for protons, the Coulomb
potential. Experimental and theoretical studies of exotic light nuclei with
a normal, localized nuclear core and a dilute few-neutron halo or skin are
now well advanced and are becoming increasingly sophisticated. Such systems
provide a stringent test of nuclear structure models developed for stable
nuclei as they involve new structures and surface phenomena.

The question about the halo nature of $^{19}C$ and its underlying structure
is one of the interesting current questions in dripline physics. The neutron
rich $^{19}C$ nucleus, the last particle-stable odd-neutron isotope of
carbon, has drawn much attention recently as candidate for having a
one-neutron halo structure \cite{Bazin}-\cite{Banerjee}. This can be easily
anticipated due to its very small neutron binding energy (of the order of a
few hundred $keV$). However, halo formation not only depends upon the small
separation energy of the valence neutron. The spin-parity and the
configuration of the single particle state of the valence neutron also play
important roles in the formation of the halo. The case in which $^{19}C$ has
a ground state $J^\pi =1/2^{+}$ with a dominant $^{18}C$ ($0^{+}$)$\otimes
1s_{1/2}$ single particle configuration \cite{Warburton} favours the halo
formation. On the other hand, the case of a $J^\pi =5/2^{+}$, expected from
standard shell model ordering, prevents the halo from being formed due to
the large centrifugal barrier associated with a $0d_{5/2}$ orbit. The $J^\pi
=5/2^{+}$ may also result when the dominant single particle configuration in
the ground state is $^{18}C$ $(2^{+})\otimes 1s_{1/2}$. In this case, the
binding energy of the valence neutron is effectively increased by the
excitation energy of $1.62$ $MeV$ of the excited ($2^{+}$) state of the $%
^{18}C$ core, thereby also hindering the halo formation. A $J^\pi =3/2^{+}$
is also possible with this configuration for the ground state, as considered
in the recent model calculations \cite{Ridikas}.

In order to probe the structure of $^{19}C$, only a few experiments on the
dissociation of this nucleus have been carried out so far \cite{Bazin},\cite
{Marques},\cite{Baumann}-\cite{Nakamura}. Very recently the relative energy
spectrum of $^{19}C$ has been measured in its Coulomb dissociation at $67$ $%
MeV/nucleon$ incident energy on $Pb$ target in a kinematically complete
experiment at Riken \cite{Nakamura}. The Riken data, when compared with
semi-classical calculations \cite{Nakamura}, support a dominant $1s_{1/2}$
configuration of $^{19}C$ halo wavefunction. The MSU data \cite{Bazin}, on
the other hand, suggest an $s-$wave neutron around the $2^{+}$ excited state
of $^{18}C$. Recent theoretical calculations on the Coulomb breakup of $%
^{19}C$ \cite{Banerjee} indicate the possibility of an $s-$wave neutron
coupled to the $2^{+}$ excited state of $^{18}C$, or a $d-$wave neutron, or
a coherent superposition of these two configurations as far as the recent
Ganil data \cite{Liegard} are concerned. The total one-neutron removal cross
sections in the Coulomb dissociation of $^{19}C$ on heavy targets have been
calculated in \cite{Banerjee}. These cross sections, when compared with
experimental data one-neutron removal cross sections measured at Ganil \cite
{Liegard} and MSU \cite{Bazin}, show that the configuration of a $1s_{1/2}$
neutron coupled to the $0^{+}$ $^{18}C$ core has non-significant
contribution to the ground state structure of $^{19}C$ \cite{Banerjee}. In
view of all these discrepancies, it has not been possible to draw any
definite conclusion on the structure of $^{19}C$.

The difficulties in drawing conclusion about the structure of $^{19}C$ might
arise also due to lack of precise knowledge of the one-neutron seperation
energy $S_n$ in $^{19}C$. This quantity is very crucial as far as the
property of the neutron halo is concerned: the asymptotic behaviour of the
radial wavefunction is governed by the separation energy and the extension
of the halo is therefore directly linked this quantitiy. The value used in
most of the calculations to date is $240$ $keV$. This value is a world wide
average of several experiments each with significant uncertainties \cite
{Vierira}. The recent measurement of the relative energy spectrum of $^{19}C$
in its breakup on $Pb$ at Riken has led to an indirect determination of the $%
^{19}C$ one-neutron separation energy \cite{Nakamura}. A semi-classical
analysis of this new, kinematically complete, measurements of the Coulomb
dissociation of $^{19}C$ showed the data to be consistent with $S_n\approx
530$ $keV$. In the present calculations, we will consider the consistency of
this proposal with the interaction cross sections.

In this brief report, we have re-investigated the structure of $^{19}C$
through the study of pickup reactions involving $^{19}C$ nucleus at low
energies, as recently the use of low energy single nucleon transfer
reactions for structure studies of exotic nuclei have attracted attention
\cite{Winfield},\cite{Lenske}. Because of the simplicity of the theoretical
interpretation of such reactions, they are thought to provide an important
source of information about the structure of halo nuclei.

Single nucleon transfer reactions, such as the $(p,d)$ and $(d,p)$
reactions, have been a reliable tool in nuclear spectroscopic studies of
stable nuclei. Besides investigating the energy level structure of such
nuclei, the spins and parity $(J^\pi )$ of levels can be determined from the
shape of the differential cross section angular distribution. The shape of
an angular distribution of an unknown level is either compared to those of
known levels (ideally in the same nucleus), or to Distorted Wave Born
Approximation (DWBA) calculations. By this technique, at least the
transferred orbital angular momentum $(\ell _n)$ may be determined, and for $%
J^\pi $ levels with relatively simple structure, one can usually then deduce
$J^\pi $ from shell model arguments.

Since radioactive beams of sufficient intensity are now starting to become
available, low energy single nucleon transfer reactions are again being
proposed as a spectroscopic tool, this time in inverse kinematics, since $%
\beta -$unstable nuclei are produced as secondary beams, with a proton or
deuteron target for the radioactive beam to investigate the structure of
nuclei far off stability. Since single nucleon transfer reactions are binary
processes the interpretation of the results is comparatively simple and
decisive answers on nuclear structure and also insight into the reaction
dynamics of exotic nuclei can be expected. The discussion in this report is
limited to the $(p,d)$ reaction. A more general review of light-ion transfer
reactions in inverse kinematics may be found in Ref. \cite{Hardy}.

It would thus be of considerable interest to measure the spin and parity of $%
^{19}C$ via the $(p,d)$ reaction. The shape of the angular distribution of
the $^{19}C(p,d)^{18}C$ reaction, measured in inverse kinematics, might be
used to distiguish between $\ell _n=0$ and $\ell _n=2$ transfer, if the last
neutron is in a pure single particle state. Unfortunately, to the best of
our knowledge, the beam intensity of $^{19}C$ produced by present facilities
put this out of reach for the moments. Nevertheless, the results of the
present work would be beneficial for the experimentalists in the near future.

A measurement of the angular distribution, hovewer, for the $%
^{17}C(p,d)^{16}C$ reaction is more feasible. However, like $^{19}C$, the
ground state spin and parity for $^{17}C$ is not known experimentally.
Again, the energies of the $5/2^{+}$, $3/2^{+}$, and $1/2^{+}$ levels given
by the shell model are too close to constitute a prediction, although a
comparison of the experimental and theoretical beta-decay lifetimes favours $%
3/2^{+}$or $5/2^{+}$. Knowledge of the $^{17}C$ ground-state spin would help
determine the shell model interaction in this $(A,Z)$ region, which would
lead to better confidence in the predictions for $^{19}C$. Hence, we start
first with the analysis of the $^{17}C(p,d)^{16}C$ reaction before
investigating the cross section angular distributions of the $%
^{19}C(p,d)^{18}C$ reaction for different possible halo-neutron
configurations.

\section{Application}

The most dramatic feature of halo nuclei is their very loosely bound
few-body character, with a strong spatial localization of the core nucleons
and a delocalization of the halo particle(s). More recent theoretical
analyses have shown that an explicit treatment of this correlated few-body
nature, or granularity, is important quantitatively for calculations of
reaction cross sections \cite{Al-Khalili}. Such a few-body description leads
to larger/smaller calculated reaction cross sections depending upon the
scattering angle than those obtained from the standard calculations which do
not consider the loosely bound structure. Such description has significant
implications for the deduced size and ground state structure of the halo. It
follows that interaction cross section analyses are model dependent and
require the use of theoretical relative motion wavefunctions for these
few-body structures.

We discuss the exit channel of the $^{17}C(p,d)^{16}C$ and $%
^{19}C(p,d)^{18}C $ reactions within three-body (halo neutron $(n)+p+core$
nucleus) models. We calculate the transfer amplitude using the prior form of
the $(p,d)$ matrix element, thus the transition interaction is the $n-p$
interaction and we need a full (three-body) description of the $n+p+core$
system in the final state. For the description of this final state we have
used the Adiabatic (AD) model \cite{Johnson} and the Quasi-adiabatic (QAD)
approach \cite{Amakawa}. To clarify the importance of the breakup
corrections to the transfer cross sections, we also perform the standard
DWBA which do not account for the effects arising from the breakup of the
projectile in the field of the nucleus.

In the AD treatment it is assumed that the excitation of the projectile is
to states in the low energy continuum; its treatment of possible high energy
breakup contributions is therefore naturally suspect. An improved treatment
of these higher energy breakup configurations is provided by the QAD method
calculations which takes approximate account of modifications to the centre
of mass energy of the $n-p$ pair in breakup configurations through the use
of a mean breakup relative energy for the continuum states. It thus breaks
the degeneracy with the elastic channel. The inclusion of these higher
energy breakup components leads, in general, to an enhanced transition
amplitude and a better description of experimental data \cite{Amakawa}.
These quantum mechanical three-body models have been well described in Ref.
\cite{Lenske} by the work of Yilmaz and G\"{o}n\"{u}l through a successful
application of the models in analyzing a halo transfer $^{11}Be(p,d)^{10}Be$
reaction. The details of the resulting partial wave expansions and solution
of the equations can also be found in Sec. III of Ref. \cite{Amakawa}.

The calculated two-body (DWBA) wavefunction, using deuteron folded
potential, and three-body (AD, QAD) wavefunctions using the same inputs, in
partial wave form, are employed in a modified version of the program TWOFNR
\cite{Igarashi} for the evaluation of the reaction observables, performing
the zero-range approximation. The radial integrals are carried out from $0$
to $40$ $fm$ in steps of $0.1$ $fm$. The maximum number of partial waves
used is $30$ for both entrance and exit channels. As the long tail of the
halo wavefunction makes internal contributions less important, all
calculations presented in this paper are done without non-locality
corrections.

\subsection{The $^{17}C(p,d)^{16}C$ Reaction at $30$ $MeV$}

As an additional test, we first deal with the ground state of $^{17}C$. The
separation energy of the last neutron in this nucleus is again small, only $%
S_n\approx 730$ $keV$ \cite{Audi}. Data concerning the structure of the $%
^{17}C$ nucleus is quite poor and can easily be summarized in a few
sentences \cite{Tilley}. Observation of beta-delayed neutron emission has
been reported \cite{Reeder}. An excited state of $^{17}C$ is observed at $%
295\pm 20$ $keV$. Three closely spaced low-lying states are expected again
like for the $^{19}C$ case, i.e. $J^\pi =1/2^{+}$, $3/2^{+}$, $5/2^{+}$, and
it is not clear which one of them is the ground state. Nevertheless, based
on the systematics described in \cite{Warburton2}, where a modified
Millener-Kurath interaction is employed within the shell model formalism, it
is unlikely that $^{17}C$ has a ground state $J^\pi =5/2^{+}$. Morever,
analysis of the gamma-ray spectra following the beta-decay of $^{17}C$ does
not support $5/2^{+}$ as the ground state spin either \cite{Warburton2}.
Unfortunately, as is clearly stated in \cite{Warburton2}, the present shell
model calculations cannot give a definitive prediction for the ground state
spin of $^{17}C$, and $1/2^{+}$ and $3/2^{+}$ both remain viable
possibilities. On the other hand, both shell model \cite{Warburton} and
relativistic mean field calculations \cite{Zhongzhou} argue for a $J^\pi
=3/2^{+}$ non-halo ground state structure for this case. In a recent paper
\cite{Herndl}, neutron capture reaction rates on neutron-rich $C-,$ $N-$,
and $O-$isotopes have been calculated in the framework of a hybrid compound
and direct capture model. The results have been compared with those of
previous calculations as well as experimental results. In that work, Herndl
{\it et al }have obtained the neutron spectroscopic factors for the possible
ground states of $^{17}C$. We make use of these spectroscopic factors in the
present calculations discussed in this section for more realistic absolute
values of the cross sections.

To investigate the difference in shapes of the angular distributions with
the various possible ground state spin assignments, we have performed
calculations with the zero-range approximation. The binding potentials in
all cases are taken to be of Woods-Saxon type with the radius and
diffuseness parameters as $1.25$ $fm$ and $0.65$ $fm$ respectively. Their
depths have been calculated to reproduce the corresponding binding energies.
The entrance and exit channel potentials, including spin-orbit interactions,
are obtained from the global parameterization of Bechetti and Greenless \cite
{Bechetti}.

We have found that the shapes of the angular distribution are not sensitive
to the radius of the neutron binding potential. Test calculations involving
different sets of potential parameters have shown that the exact location
and sharpness of extrema in the cross sections depend on the choice of
optical model potential but there is always a clear distinction between $%
\ell _n-$transfers. In particular, the $\ell _n=0$ cross section always has
a maximum at $\theta _{c.m.}=0^{\circ }$ and falls off rapidly with angle.
From the calculations shown in Figs. \ref{Fig.1.} and \ref{Fig.2.}, it is
seen that at $30$ $MeV/nucleon$ one would require to measure cross sections
out to at least $\theta _{c.m.}=15^{\circ }$ to distinguish clearly between $%
\ell _n=0$ and $\ell _n=2$ transfer. Despite the uncertainties on optical
potentials which in fact are important for a precise description of transfer
reaction cross sections, the results can be expected to describe
realistically the dynamical features of such reactions. Thus, such a
measurement, besides being of interest in its own right, would constitute a
first step towards a future $^{19}C$ experiment.

In addition, Fig. \ref{Fig.1.} makes clear that the inclusion of breakup
effects in halo transfer reactions produces changes in the shape of the
angular distributions and these effects increase the absolute values of the
theoretical cross sections in forward direction and thus lead to the smaller
values of the spectroscopic factors extracted from the experimental data. It
is expected that such effects will be larger when the incident energy
increases and the mass of the target decreases. We thus will only consider
the more exact QAD model calculation results in the next section in treating
the deuteron channel of the $^{19}C(p,d)^{18}C$ reaction.

\subsection{The $^{19}C(p,d)^{18}C$ Reaction At $30$ $MeV$}

Since the spin/parity for the ground state of $^{19}C$ is still
experimentally unknown, we have considered in our calculations the following
configurations for the valence neutron in $^{19}C$:

\begin{description}
\item[(a)]  a $1s_{1/2}$ state bound to a $0^{+\text{ }18}C$ core by $0.24$ $%
MeV$,

\item[(b)]  a $1s_{1/2}$ state bound to a $2^{+\text{ }18}C$ core by $1.86$ $%
MeV$, \cite{Tilley2}

\item[(c)]  a $0d_{5/2}$ state bound to a $0^{+\text{ }18}C$ core by $0.24$ $%
MeV$, and

\item[(d)]  a $1s_{1/2}$ state bound to a $0^{+\text{ }18}C$ core by $0.53$ $%
MeV$.
\end{description}

Since the separation energy $S_n$ is not known with satisfactory accuracy it
is treated as an input parameter in our calculations. Therefore we have also
performed calculations {\em (d)} for a larger value, $S_n=530$ $keV$, to
investigate the sensitivity of the results to the separation energy.

The parameterization procedure of the $^{18}C$ core and the single neutron
interaction within the QAD model is again taken similar to the case of the
previuos calculations. We initially, see Fig. \ref{Fig.3.}, assume a
Woods-Saxon $n+^{\text{ }18}C$ binding interaction with radius parameter $%
r_0=1.25$ $fm$, diffuseness $a_0=0.65$ $fm$, and neutron separation energies
$0.24$ $MeV$ and $1.86$ $MeV$ for the core ground and excited state
configurations, respectively. However, in order to consider the cross
section sensitivity to the assumed neutron binding potential geometry $%
(r_0,a_0)$, we perform our calculations with different sets of potential
parameters. We see that the cross sections at large angles increase with
increasing $a_0$ and $r_0$ parameters while they seem similar at forward
angles.

As seen from Fig. \ref{Fig.3.}, reducing $S_n$ does not considerably change
the shape and magnitude of the cross section angular distributions for the
reaction considered, unlike the case {\em (b)} that leads to the larger
cross sections with a different absolute value at forward angles when
compared to the case of {\em (a).}

In Fig. \ref{Fig.4.}, the energy dependence of the cross section for a
possible neutron-halo configuration of the reaction under study is shown.
This observation might be of interest for the design of experimental setups.
The small separation energy and large spatial extension of the neutron-halo
wavefunction have lead to an energy dependence of the cross sections which
is very different from these known for transfer reactions on well bound
systems. The rapid decrease of the calculated cross sections with increasing
incedent energy for the weakly bound nucleus shows that the dynamics of
transfer reactions is favorable for systems far off stability and
experiments can very likely take advantange of this behaviour. Only transfer
reactions with rather large cross sections can be investigated due to the
low intensity of secondary beams and the use of inverse kinematics in
reactions involving halo nuclei. It is clear from Fig. \ref{Fig.4.} that low
energy transfer reactions indeed might be useful for investigations of such
short-lived isotopes.

\section{Summary and Conclusion}

Recently experimental evidence for a new one-neutron halo candidate, namely $%
^{19}C$, has become available. This has sparked off considerable interest as
the only established one-neutron halo nucleus so far has been $^{11}Be$
where relative $s-$motion dominates the ground state. The $^{11}Be$ nucleus
has been extensively studied during the last few years and has provided a
testing ground for single neutron halo theories [12, and the references
therein].

The structure of the $^{19}C$ ground state and its last neutron separation
energy $S_n$ are however still very uncertain. A naive shell model suggests
a nodeless $0d_{5/2}$ orbit for the least bound neutron. More detailed
calculations \cite{Warburton} however predict a $1/2^{+}$ $^{19}C$ ground
state due to a lowering of the $1s_{1/2}$ orbital. A $3/2^{+}$ or $5/2^{+}$
ground state remains a theoretical possibility \cite{Ridikas}, but would
involve $(2^{+}\otimes 1s_{1/2})$ and/or $(2^{+}\otimes 0d_{5/2})$ excited $%
^{18}C$ core components, and $(2^{+}\otimes 1s_{1/2})$ and/or $(0^{+}\otimes
0d_{5/2})$ components, respectively.

To investigate this uncertainty, we have taken into account the idea that in
combination with independent experimental data, such as from Coulomb
dissociation measurements and from momentum distributions following $^{19}C$
breakup reactions, cross section angular distribution measurements for halo
transfer reactions involving $^{19}C$ nucleus may at this stage provide a
useful constraint on the ground state of such nuclei. Along this line, we
have studied the $^{19}C(p,d)^{18}C$ neutron-halo transfer reaction,
together with the test calculations on the $^{17}C(p,d)^{16}C$ reaction,
performing the calculations for both $1s_{1/2}$ and $0d_{5/2}$ neutron
configurations with separation energy $S_n=0.24$ $MeV$ about a $%
^{18}C(0^{+};g.s)$ core, and with separation energy $S_n=1.86$ $MeV$ about a
$^{18}C(2^{+};1.62$ $MeV)$ excited core. The comparison between the
calculation results, which have been obtained using approximate two- and
three-body quantum mechanical models for different possible single neutron
configurations, gives some support to the suggestion that at $30$ $%
MeV/nucleon$ incident energy one would require to measure cross sections out
to at least $\theta _{c.m.}=15_{\text{ }}^{\circ }$ to distinguish clearly
between $\ell _n=0$ and $\ell _n=2$ transfer for the reaction of interest.

The $(p,d)$ reaction is a potentially useful spectroscopic tool in inverse
kinematics with radioactive beams. Its advantages include its simplicity
(reliability of model calculations) and relatively large cross sections for
single particle states. A promising aspect for the future is that the
instrumental energy and angular resolution requirements will be more
relaxed. The high intensities of radiaoactive beams will not only allow
transfer reactions to be performed with exotic isotopes, but will also
permit collimation of beams close to the stability line, thus giving better
emittance quality.

\newpage\

{\bf FIGURE CAPTIONS}

Fig. \ref{Fig.1.} Calculated differential cross section angular
distributions obtained by different theoretical models for the reaction $%
^{17}C(p,d)^{16}C$ at $30$ $MeV$ for a $1s_{1/2}$ , $0d_{3/2}$ and $0d_{5/2}$
transfer. The spectroscopic factors are set to 0.644, 0.035 and 0.701,
respectively.

Fig. \ref{Fig.2.} Calculated angular distributions using the QAD model for
the transitions as in Fig. \ref{Fig.1.}

Fig. \ref{Fig.3.} Calculated differential cross section angular
distributions obtained by the QAD model for the $^{19}C(p,d)^{18}$ reaction
at $30$ $MeV$ for possible neutron-halo configurations. The spectroscopic
factor is 1.

Fig. \ref{Fig.4.} As for Fig. \ref{Fig.3.}, but the QAD calculations have
been carried out for different bombarding energies in case of a $%
(0^{+}\otimes 1s_{1/2})$ neutron-halo configuration.

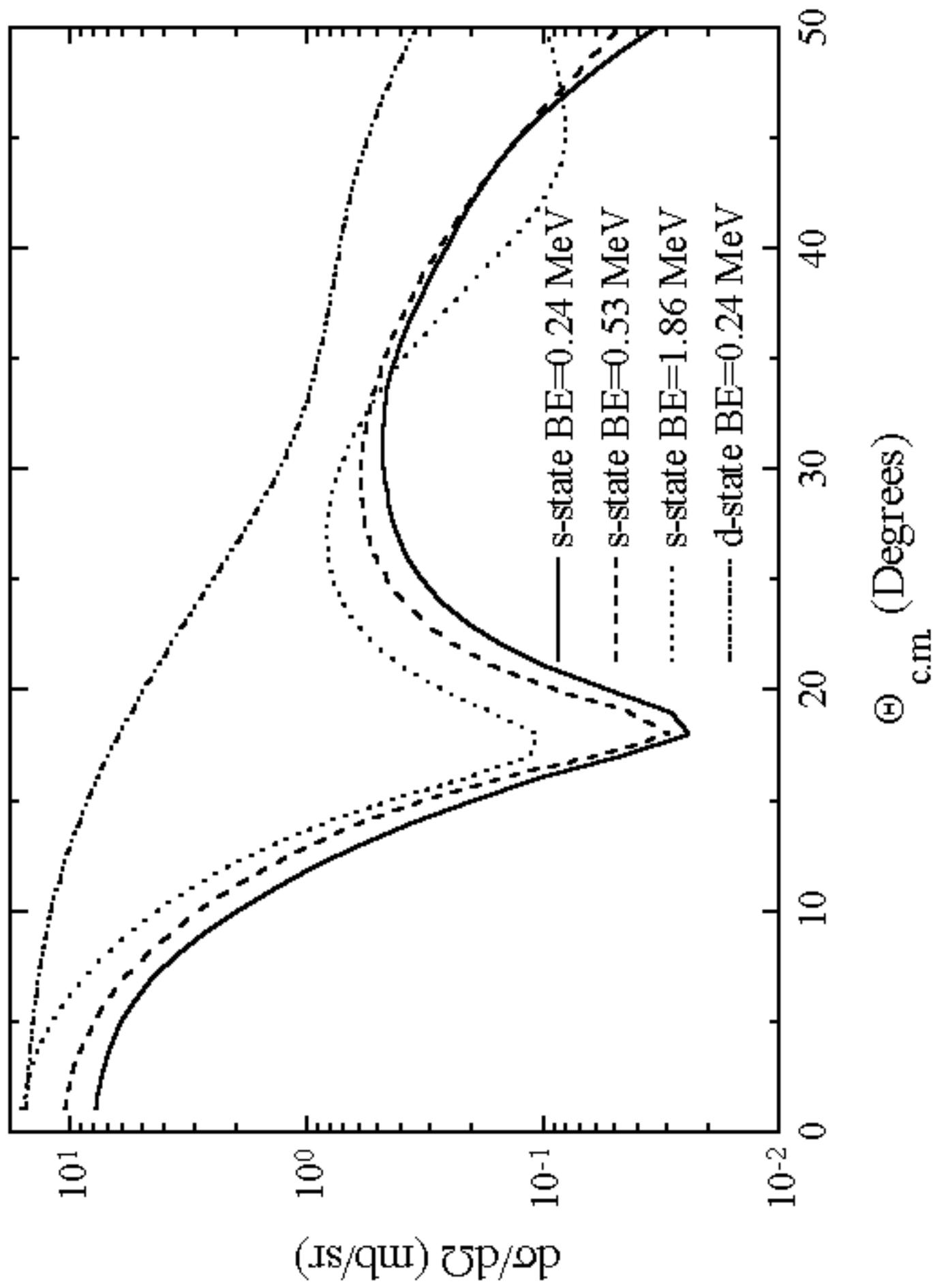
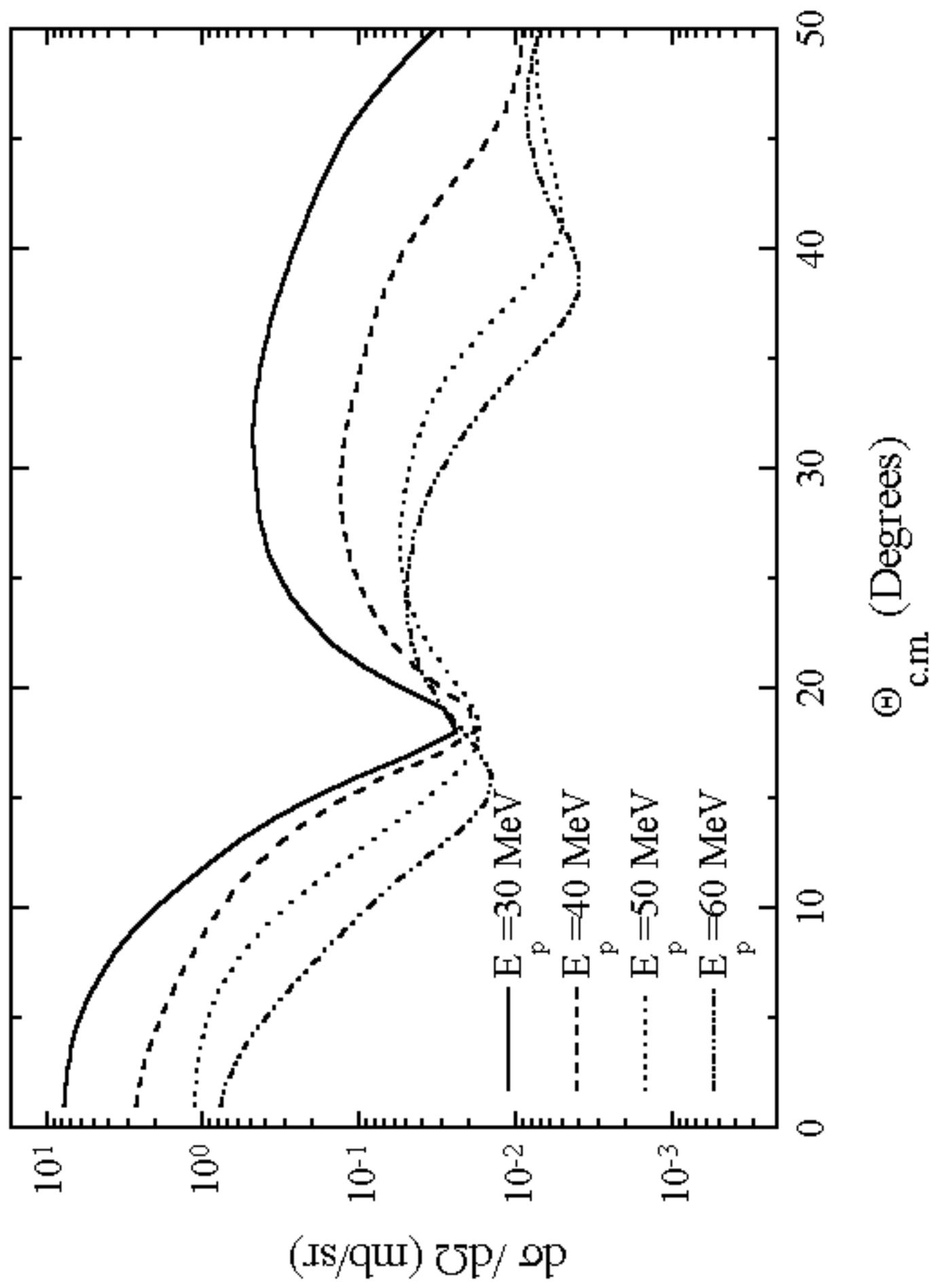
\end{document}